\documentclass[fleqn,10pt]{wlscirep}
\usepackage{xspace}
\usepackage[utf8]{inputenc}
\usepackage[T1]{fontenc}
\usepackage{todonotes}

\usepackage{url}

\newcommand{\R}{$\textrm{R}_0$\xspace}

\newcommand{\Reff}{$\textrm{R}_\textrm{eff}$\xspace}

\begin{document}

\title{Stochasticity and heterogeneity in the transmission dynamics of SARS-CoV-2}

\author[1,2,3,*,+]{Benjamin M. Althouse}
\author[1,+]{Edward A. Wenger}
\author[4]{Joel C. Miller}
\author[5,6,7,8,9]{Samuel V. Scarpino}
\author[10,11]{Antoine Allard}
\author[10,12,13]{Laurent H\'{e}bert-Dufresne}
\author[14,+]{Hao Hu}


\affil[1]{Institute for Disease Modeling, Bellevue, WA}
\affil[2]{University of Washington, Seattle, WA}
\affil[3]{New Mexico State University, Las Cruces, NM}
\affil[4]{School of Engineering and Mathematical Sciences, La Trobe University, Bundoora, VIC, Australia}
\affil[5]{Network Science Institute, Northeastern University, Boston, MA}
\affil[6]{Department of Marine \& Environmental Sciences, Northeastern University, Boston, MA}
\affil[7]{Department of Physics, Northeastern University, Boston, MA}
\affil[8]{Department of Health Sciences, Northeastern University, Boston, MA}
\affil[9]{ISI Foundation, Turin, Italy}
\affil[10]{D\'epartement de physique, de g\'enie physique et d'optique, Universit\'e Laval, Qu\'ebec (Qu\'ebec), Canada}
\affil[11]{Centre interdisciplinaire en mod\'elisation math\'ematique, Universit\'e Laval, Qu\'ebec (Qu\'ebec), Canada}
\affil[12]{Vermont Complex Systems Center, University of Vermont, Burlington, VT}
\affil[13]{Department of Computer Science, University of Vermont, Burlington, VT}
\affil[14]{Bill \& Melinda Gates Foundation, Seattle, WA}

\affil[*]{balthouse@idmod.org}

\affil[+]{these authors contributed equally to this work}


\begin{abstract}

SARS-CoV-2 causing COVID-19 disease has moved rapidly around the globe, infecting millions and killing hundreds of thousands. The basic reproduction number, which has been widely used and misused to characterize the transmissibility of the virus, hides the fact that transmission is stochastic, is dominated by a small number of individuals, and is driven by super-spreading events (SSEs). The distinct transmission features, such as high stochasticity under low prevalence, and the central role played by SSEs on transmission dynamics, should not be overlooked. Many explosive SSEs have occurred in indoor settings stoking the pandemic and shaping its spread, such as long-term care facilities, prisons, meat-packing plants, fish factories, cruise ships, family gatherings, parties and night clubs. These SSEs demonstrate the urgent need to understand routes of transmission, while posing an opportunity that outbreak can be effectively contained with targeted interventions to eliminate SSEs. Here, we describe the potential types of SSEs, how they influence transmission, and give recommendations for control of SARS-CoV-2.

\end{abstract}

\flushbottom
\maketitle

\thispagestyle{empty}


\noindent Corresponding author:\\
Benjamin M Althouse\\
Institute for Disease Modeling\\
3150 139th Ave SE\\
Bellevue, WA, 98005\\
Phone: (425) 777-9615\\
Email: balthouse@idmod.org\\

\clearpage
\pagebreak

\section*{Significance statement}

SARS-CoV-2 causing COVID-19 disease has moved rapidly around the globe, infecting millions and killing hundreds of thousands. The basic reproduction number, which has been widely used and misused to characterize the transmissibility of the virus, hides the fact that transmission is stochastic, is dominated by a small number of individuals, and is driven by super-spreading events (SSEs). Many explosive SSEs have occurred in indoor settings such as long-term care facilities, prisons, meat-packing plants, fish factories, cruise ships, family gatherings, parties and night clubs. These SSEs demonstrate the urgent need to understand routes of transmission, while posing an opportunity that outbreak can be effectively contained with targeted interventions to eliminate SSEs. 

\section*{Introduction}

While Severe Acute Respiratory Syndrome coronavirus 2 (SARS-CoV-2) has moved swiftly around the globe, causing millions of Coronavirus Disease 2019 (COVID-19) cases, much attention has been given to the basic reproduction number (\R), estimated to be roughly between 1.5 and 4 \cite{li2020early}. As the virus has spread it has become clear that relying on a single value to characterize the number of secondary infections --- and thus estimates of the transmissibility of this virus --- is inadequate to capture the true transmission dynamics and subsequent risk to humanity~\cite{hebert-dufresne2020r0a}. Indeed, a litany of official reports and anecdotes have identified key superspreading events (SSEs) which have propelled transmission and infected many.

Earliest notable examples include a Briton who returned from a Singapore conference and infected 13 other people in a ski resort in the Alps~\cite{hodcroft2020preliminary}; more than 70 cases were linked to a Boston Biogen conference within 2 weeks~\cite{biogen}; and the most extreme example so far, South Korea ``Patient 31'' started a cluster with more than 5,000 cases in Daegu~\cite{reuters.seoul}. A Hong Kong resident visited the Diamond Princess cruise ship on January 25 and later tested positive. The number of positive cases on the cruise quickly rose to about 700 people, or 17\% of all passengers within 20 days ~\cite{mizumoto2020estimating}. In Chicago, 15 cases stemmed from one person at multiple family gatherings~\cite{ghinai2020community}. An average of 35\% secondary attack rate was found in multiple reported clusters of 2 to 48 people having meals together~\cite{liu2020secondary}.

In March, multiple European countries simultaneously reported unusually large numbers of imported cases from Ischgl, Austria, a popular ski town. Epidemiological investigations found infections might have been circulating since late February, and the individual causing the SSEs might have been a bartender working in an apr\`es ski bar who subsequently was diagnosed positive~\cite{correa-martinez2020pandemic}. In New York, a lawyer was infected and spread to at least 50 others in New Rochelle~\cite{cnn.lawyer}. In Mount Vernon, WA after a choir rehearsal on March 10th, 45 out of 60 choir members fell ill, and 28 of those 45 tested positive for SARS-CoV-2. No one appeared to be sick during the rehearsal~\cite{komo.choir}. Investigations revealed that the index patient in this case directly infected 52 others, with an the attack rate of 53.3\% to 86.7\%~\cite{hamner2020high}. An Indian preacher died after returning from a trip to Italy and Germany, and attending a large gathering to celebrate the Sikh Festival of Hola Mohalla. A week later, at least 19 of his relatives were infected, and it resulted in a quarantine of 40,000 residents in Punjab~\cite{bbc.india}.

In April and May, universal testing at a Boston homeless shelter found 36\% of residents were tested PCR positive~\cite{baggett2020covid19}, and an 87-year-old in Harbin, China directly infected more than 78 individuals within a few days at home and two hospitals~\cite{dw.china}. Singapore has seen a sharp rise in cases, with the vast majority (88\%) being linked to dormitories. S11, a 10,000 person capacity dormitory, has the largest cluster in Singapore, with more than 2,200 infections~\cite{nyt.dorm}. More than 100 cases were traced back to nightclubs in Seoul that were visited by a young man who was later tested positive~\cite{seoul.nightclub}. In Chennai, India, the Koyambedu vegetable market has emerged as 
a hotspot, more than 2000 cases have been traced to the market~\cite{india.market}.

These accounts in addition to the many examples of SSEs in long-term care facilities~\cite{mcmichael2020epidemiology}, prisons~\cite{akiyama2020flattening,kinner2020prisons}, meat processing facilities~\cite{dyal2020covid}, and fish factories~\cite{ghanafish} demonstrate the central role played by SSEs on the transmission dynamics -- and subsequent epidemiological control of -- SARS-CoV-2. Here, we describe the potential types of SSEs, how they influence transmission, and give recommendations for control of SARS-CoV-2.

\section*{Superspreading events, stochasticity, and negative binomially-distributed secondary infections}

\begin{figure}[t!!]
\centering
\includegraphics[width=\linewidth]{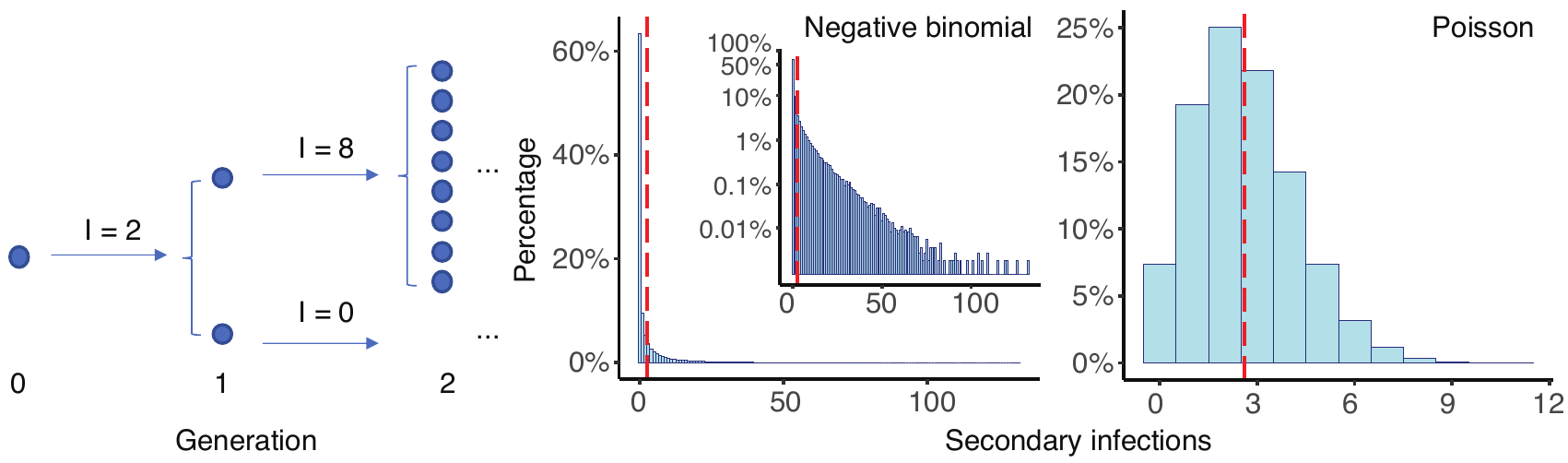}
\caption{\textbf{Poisson and negative binomially-distributed (NB) secondary infections.} Figure shows a schematic of a NB branching process, and example distributions of NB and Poisson distributions of secondary infections. Both distributions have \R = 2.6~\cite{defaultR0}and the NB has dispersion parameter $k=0.16$~\cite{endo2020estimating} with 100,000 draws.}
\label{fig:schematic}
\end{figure}

Many classical models in epidemiology either assume or result in a Poisson distribution of secondary infections per infected individual.  Because Poisson distributions have the same mean and variance, they often fail to capture the relevant features of SSEs.  As a result, SSEs are now commonly modeled using a negative binomial (NB) distribution of secondary infections per infected individual~\cite{lloyd-smith2005superspreading}. NB distributions can be parameterized with a mean (thought of as \R) and a dispersion parameter, $k$, where smaller values of $k$ indicate a longer tail (over-dispersion). When $k$ is close to zero, even with a high \R, most individuals give rise to zero or one secondary infections, and few give rise to many infections -- a so-called `long tail' of infections. However, as $k$ grows larger, a NB distribution approaches a Poisson distribution (becoming exactly Poisson when $k \to \infty$) and the effect of SSEs on the epidemic decreases. In Figure~\ref{fig:schematic} we show a schematic of a NB branching process, and distribution of NB and Poisson distributions of secondary infections under the same \R. Earlier transmission modeling studies suggest the offspring distribution of SARS-CoV-2, the etiological agent of COVID-19, is highly over-dispersed with $k$ comparable to SARS-CoV-1~\cite{endo2020estimating}.

\begin{figure}[t!!]
\centering
\includegraphics[width=\linewidth]{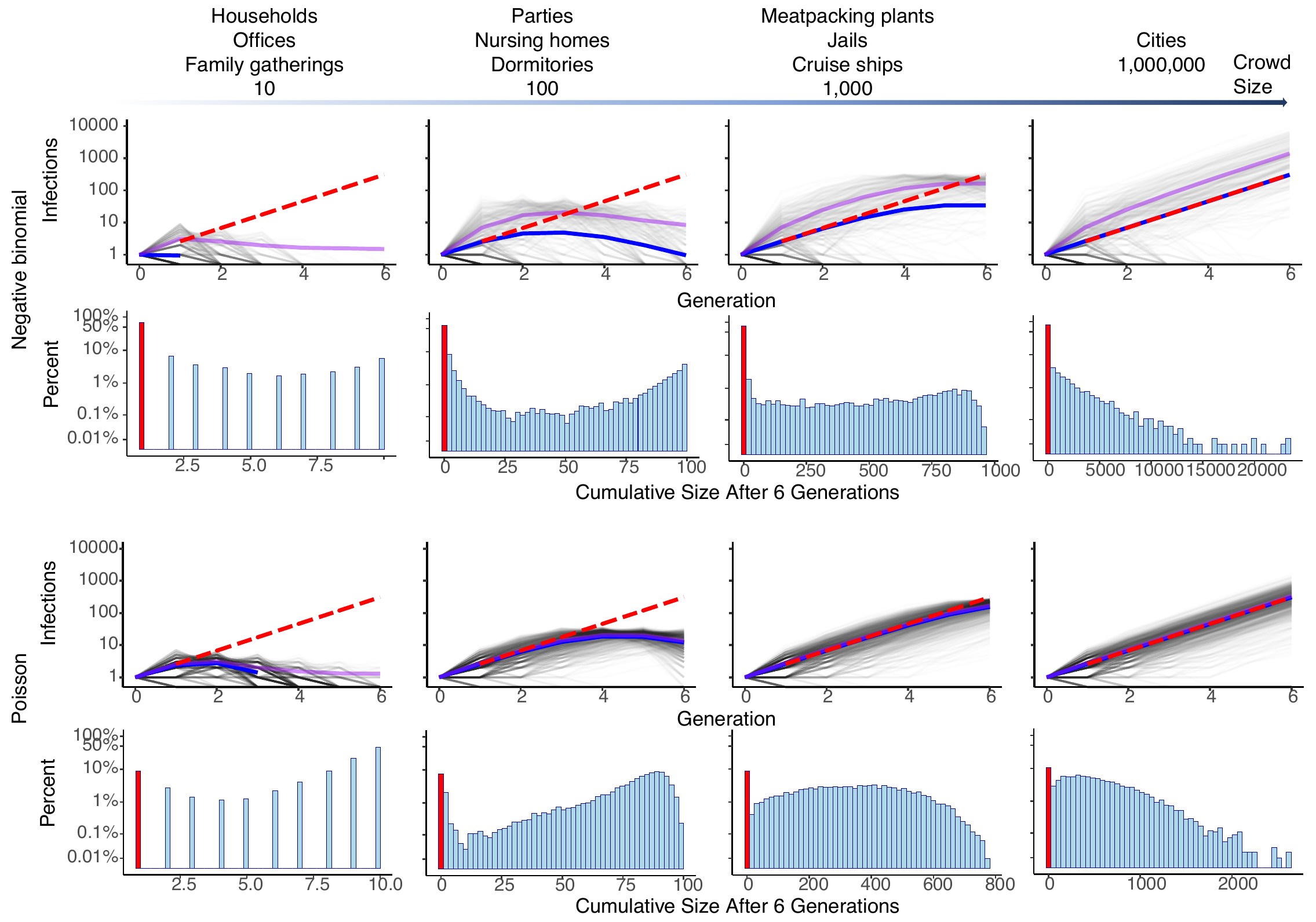}
\caption{\textbf{Example trajectories of NB and Poisson branching processes.} Figure shows example trajectories (in number of active infections vs generation) of NB and Poisson branching processes, and cumulative infection sizes after 6 generations of spread. Both simulations start with 1 infection, have the same \R = 2.6. For NB branching process we assume dispersion parameter $k$ = 0.16, same as SARS-CoV-1~\cite{lloyd-smith2005superspreading, hellewell2020feasibility}. We run all simulations 10,000 times. Dashed red lines represent theoretical values $I = R_0^n$ where $I$ is number of active infections and $n$ is number of generations. Solid blue lines are the mean values of all simulations including those that have not taken off, which overlap with the theoretical values when the susceptibles are not depleted. Solid purple lines are the mean value for simulations that took off, and the outbreaks appear more explosive in the first few generations in the NB simulations. Both number of active cases and cumulative infections are in $\log_{10}$ scale.}
\label{fig:trajectories}
\end{figure}

An outbreak dominated by SSEs and a NB distribution of secondary infections with small $k$ has distinct transmission features as compared to a Poisson distributed, `mean-field' outbreak with the same \R. First, secondary infections will be overdispersed, such that early dynamics are more stochastic and it is less likely that an outbreak will grow large. In Figure~\ref{fig:trajectories} we show an example by utilizing a stochastic branching process model with both Poisson and SARS-CoV-1 like NB distribution ($k=0.16$) under the same mean $R_0=2.6$~\cite{defaultR0}, with different population sizes ranging from small clusters of $10$ like households to large ones of $10^6$ like city-wide. We observe the difference of transmission dynamics in 6 generations -- 24-36 days using a 4-6 day serial interval~\cite{duearly, nishiura2020serial, he2020temporal,pung2020investigation}. 
In all cases where population size $\geq 100$, 63\% of outbreaks have no secondary infections, and 77\% of outbreaks have less than 10 total infections and do not establish ongoing transmission, compared with 7\% with no secondary infections and 11\% with less than 10 total infections of simulations for a Poisson model. This is due to the over-dispersion of the SARS-CoV-1 like NB distribution: as the probability of giving rise to zero or one additional case increases, onward transmission becomes less likely to establish.
Second, if the outbreak takes off and increases beyond one of two SSEs, i.e., often only a few dozen cases, the dynamics begin to show stable exponential growth, with a growth rate approaching that of a model with the same \R, but a Poisson distribution of secondary infections per infected individual, i.e., a NB with $k \gg 0$. Third, once the outbreak takes off, it will appear more explosive (represented in estimated \R based on case counts) in the first few generations when SSEs will generate the vast majority of secondary infections, making it possible to spin out large infection clusters in few numbers of generations, whereas a Poisson model cannot. An example is shown in Figure~\ref{fig:trajectories} where the outbreak appears more explosive (purple line vs blue line) if we only consider outbreaks that took off, also consider the outbreak aboard the Diamond Princess Cruise where 135 cases were seen within 5 days \cite{mizumoto2020estimating}. Fourth, and importantly, to establish and stabilize exponential growth of an outbreak, a continuous fuel of SSEs is necessary and an outbreak can be brought largely under control --  the effective reproduction number \Reff is significantly reduced when SSEs which drive transmission are eliminated. 

\section*{Characterizing super spreading events}

The total number of secondary infections from an infected individual is based on two variables: number of susceptible contacts (potential infectees) and probability of infection per contact. 
SSEs can occur from either an infectious individual having close contacts with many susceptible individuals and/or having a higher probability of transmission per contact, or it can be an opportunistic phenomena where number of contacts and/or probability of contact is high in an unusual way, like festivals, bars, or social gatherings. There are at least four types of SSEs:
\begin{itemize}
\item Biological. Individuals with a higher probability of transmitting per contact. These may be hard to identify {\em a priori}; for infections with most pathogens, pathogen loads may vary over many orders of magnitude across individuals~\cite{woolhouse1997heterogeneities, matthews2006heterogeneous}. 
For SARS-CoV-2 individual-level viral loads are dependent on time since onset, and might be associated with demographics like age and disease severity~\cite{to2020temporal}. The temporal profile of SARS-CoV-2 viral load peaks at or just before the onset of symptoms and decreases quickly to near the PCR detection threshold within a week~\cite{he2020temporal, to2020temporal, tan2020viral}. Large heterogeneities in viral load -- up to 8 $\log_{10}$ differences -- exist between individuals~\cite{jones2020analysis}.

\item Behavioral / social. Individuals causing SSEs may have a higher number of susceptible contacts per person. Numerous studies have demonstrated marked differences in individual contacts by profession and over time. 

\item High-risk facilities and places such as meat-packing plants, workers' dormitories, prisons, long term care facilities, or health care settings. The nature of interactions in these places seem to repeatedly place individuals at higher risk of acquiring and transmitting infection. Importantly, this driver of SSEs can lead to spillover into the larger community.  Controlling a broader outbreak will be very difficult if there is a focal hotspot which is continually seeding new transmission chains, as seen in other respiratory pathogens such as tuberculosis~\cite{warren2018investigating,mabud2019evaluating}.

\item `Opportunistic' scenarios. The first scenario is when larger numbers of individuals temporarily cluster, and even with an average probability of transmission per contact, people are briefly far above their ‘average’ number of susceptible contacts. The second scenario is probability of transmission per contact is temporarily increased in an unusual way, such as singing or frequent loud speaking. These two opportunistic scenarios are more frequently seen in outbreaks at night clubs, cruise ships, crowded public transportation, parties, choirs, or other mass gathering events.

\end{itemize}
In those cases where an individual's connectivity plays a role, it is generally expected that these individuals will also be disproportionately likely to become infected as well.  These may be easier to identify in advance, but the other cases may be more difficult to prevent.

The stochastic characteristics of early transmission resulting in a dichotomy of frequent extinctions with rare, but explosive outbreaks is reflected in a few examples. As of May 16, 2020, in Ohio, only 3 out of 16 jails with positive cases had large outbreaks of attack rate larger than 10\%~\cite{ohio.jail}. In King County, WA, a total of 105 long-term care facilities reported positive cases, but only 19 of them has 5 or more deaths indicating potentially large outbreaks within facilities~\cite{kc.ltc}. Among a total of 47 cruise ships reporting positive cases or COVID-19 related deaths, only a few had notable large outbreaks, i.e. Diamond Princess, Ruby Princess, Grand Princess, Celebrity Apex, Horizon, Greg Mortimer, and Costa Atlantica~\cite{cruiseship}.

In January and February, initial clusters of explosive outbreaks did not take off first at the expected dense metropolitan areas and air transit hubs (such as Seoul, Tehran, Paris, London, or Frankfurt), but instead started in neighboring small cities: Daegu in South Korea, Gangelt in Germany, Qom in Iran, and Lombardy in Europe. This can partially be explained by considering that outbreaks are much more likely to occur when seeded by an SSE, and while large population centers will have more introductions than any given smaller city, there are many small cities, and the initial large outbreaks occur in whichever locations the first SSEs occur.   
Therefore, incidence data can look more random and relevantly unpredictable than expected if we only use population density and air traffic data as sole predictors. 

Besides stochastic SSEs, there are other factors that might contribute to such patterns. For example, a recent study found epidemic intensity of COVID-19 is strongly shaped by crowding, suggesting that there might be a spatial hierarchy of multiple population clusters in large cities, and transmission might need to try getting past a few of these clusters to start a large outbreak, which is therefore more difficult~\cite{rader2020crowding}. In contrast, in small cities people are more well-mixed and outbreaks might be easier to start.

This perspective suggests that we need to respond to all identified outbreak locales under the assumption that an SSE is occurring. If we only find the ones well established, then we miss out on potentially controlling outbreaks before they get out of control. With a strong surveillance system, we may be able to find some outbreaks early enough and might underestimate their future potential if we assume the growth will be average. Those outbreaks that do not go extinct on their own also tend to create larger epidemics. Altogether, the impact of heterogeneity highlights the importance of early outbreak detection and of local interventions.

\section*{Where do super spreading events take place?}

Since the beginning of the outbreak many types of SSEs have been reported with increasing frequency. Although the data are still scarce, some patterns have started to emerge on where and under what circumstances SSEs occur: closed environments, environments with poor ventilation, crowded places, and long durations of potential exposure all correlate with emergence of SSEs. 
In Japan, looking at clusters of transmission, researchers found closed environments facilitate secondary transmission of SARS-CoV-2 and promote SSEs; the odds ratio of SSEs in closed environments was as high as 32.6 compared with those in open spaces~\cite{nishiura2020closed}.

These SSE `hotspots' are both a source of infection for the community as well as potential targets for intervention. Hotspots are emerging as seeding infection in small metro areas as large outbreaks have occurred in meat packing plants and prisons. Many clusters were linked to a wide range of mostly indoor settings, like households, public transport, hospitals, parties, bars, elderly care centers, and schools~\cite{leclerc2020settings,qian2020indoor}. 

Because they play an important role in the spread of infection, hotspots pose an opportunity for surveillance and control: focusing on facilities and activities known to sustain hotspots, such as healthcare facilities, nursing homes, prisons, meat-packing plants, homeless shelters, schools, mass gatherings, as well as those places with closed, poorly circulated environments, can provide efficient ways to identify potential SSEs before they happen, therefore, potentially reducing a substantial amount of transmission in the population.

\begin{figure}[t!!]
\centering
\includegraphics[width=\linewidth]{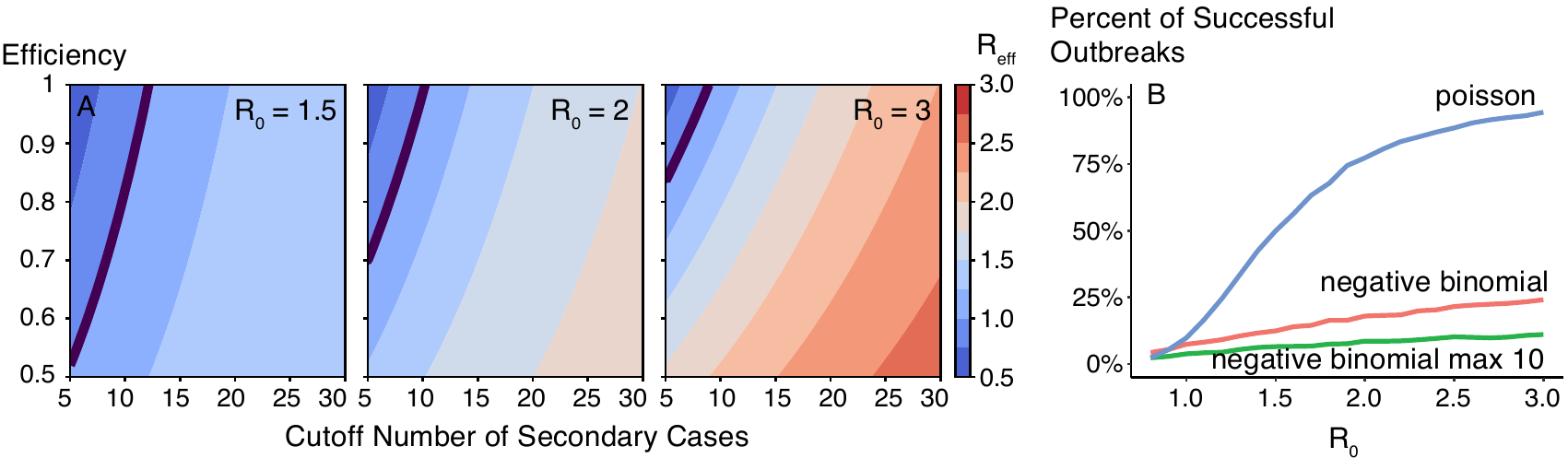}
\caption{\textbf{Controlling outbreaks and the effect of `cutting the tail'.} (A) Values of \Reff under different thresholds of maximum number of secondary infections, and the probability of keeping the maximum number of secondary infection thresholds.
We compare three scenarios with \R = 1.5, 2.0 and 3.0, where lower \R represents the consequence of population-wide mild to moderate non-pharmaceutical interventions. The NB distribution is truncated at the max number of secondary cases, and the distribution is re-normalized after the truncation. Contours highlight \Reff = 1 boundary. Lower \R facilitates the extinction of outbreak, and increases the probability of success controlling the outbreak, especially when the probability of reducing hotspot transmission is low. If the vast majority of transmission inside hotspots can be eliminated, under all \R scenarios a control target of less than 10 secondary infections can bring \Reff to close to unity. (B) Probability of an outbreak successfully taking off with one seed infection under NB and Poisson distributions, as well as a NB distribution with cutoff number of secondary infections of 10 and 100\% efficiency. The stochastic branching process simulations are run 10,000 times with \R ranging from 0.8 to 3.0. The success of outbreak is defined as having more than 20 infections at generation 6. 
Dispersion parameter $k$ for NB is 0.16. }
\label{fig:cutthetail}
\end{figure}

\section*{Controlling outbreaks with targeted interventions: 'cutting the tail'}

As seemingly rare SSEs drive the growth trajectory of an epidemic, trimming the long tail of large secondary infections is important for outbreak control and suppression (whether through targetted interventions or broad social-distancing). We demonstrate this effect in the contour plots in Figure~\ref{fig:cutthetail}(A) by randomly drawing the values of secondary infections from a NB distribution but if the value exceeds a cutoff threshold, then an efficiency parameter (from 0.5 to 1) gives the probability that the value is redrawn until it does not exceed the threshold.  Thus we assume that the behavior of most individuals does not change but we remove a fraction of the SSEs determined by our control efficiency. Using this intervention model under \R = 3, and only targeting a small number of individuals with larger than 10 secondary infections (about 10\% of all secondary infections) with 50\% to 100\% efficiency, the mean \Reff is reduced to 2.06 and 1.09, respectively.

We also present two scenarios assuming a mild to moderate non-pharmaceutical interventions (NPIs) or behavioral changes bringing population-wide \R down to 2 or 1.5, as well as with the targeted intervention on SSEs truncating number of secondary infections. With \R = 1.5, and a cutoff of $10$ secondary infections the overall \Reff is reduced to 0.85 and 1.21 under 100\% and 50\% efficiency. It appears the `cutting the tail' interventions are more effective and impactful than reducing population-wide \Reff: if probability of success is large, \Reff is under or around unity if the maximum secondary infections are truncated at 10. These results suggest stakeholders of different types of hotspots should strive to achieve the goal of controlling number of secondary infections to at least less than 10 with industry-specific mitigation efforts. This effect is also seen in Figure~\ref{fig:cutthetail}(B), looking at the probability of launching large outbreaks by branching process simulations similar as before. Results show how the probability of successfully launching a large outbreak -- defined as having more than 20 infections at generation 6 -- decreases with \Reff ranging from 0.8 to 3.0. With the same \R, by `cutting the tail' the probabilities of starting large outbreaks are much lower. These initial results suggest that the collapse of a transmission chain is happening more frequently once the long tail is gone, because there are not enough SSEs to fuel its continuous growth. 

Efforts should be spent to understand how to reduce transmission in the four types of SSEs. For the biological type, it is important to understand the characteristics of individuals causing SSEs. For example, what individual-level attributes and behaviors lead to SSEs, and how to identify them quickly? As viral loads decrease over time and most transmissions are front-loaded, timely identification and/or isolation becomes crucial.
If future studies reveal meaningful viral load differences by demographics -- for example a difference by age -- and that there exists meaningful differences in transmission across viral loads, a risk classification algorithm can be used together with individual contact networks to stratify individuals, and remind them the potential of causing SSEs as they are reported sick and need to be tested. PCR test results could potentially include a rough category of transmissibility by cycle threshold values, for example, reporting a high transmission risk for those with cycle thresholds less than 20. This might help people avoid crowded events and reduce their contacts. Environmental surveillance might play an important role in identifying the hotspots if viral load in stools from individuals causing SSEs are also orders of magnitudes greater~\cite{wu2020sars}. For the behavioral/social type, a complete contact network for an individual can be estimated using surveys or cell phone tracking and used to identify two kinds of people -- those with a high degree meaning lots of close contact per day, and those with high centrality, meaning they are crucial connecting different population clusters. Personalized messages or tests could be targeted at these individuals stressing the need for better infection control. For both high-risk facilities and `opportunistic' scenarios, to effectively prevent transmission in hotspots, it is important to understand the indoor transmission routes of SARS-CoV-2, which might include droplets and aerosols through coughing, sneezing, and speaking~\cite{meselson2020droplets,van2020aerosol,pmid32404416}, fecal-oral~\cite{hindson2020covid,cheung2020gastrointestinal,xiao2020covid} or indirectly through surfaces~\cite{ogbunugafor2020intensity}. A recent study estimates 1 minute of loud speaking generates at least 1,000 virion-containing droplet nuclei and could remain airborne for more than 8 minutes~\cite{pmid32404416}, suggesting a possible mechanism of creating large exposure and starting SSEs. Contact tracing is a useful tool to reveal the relative importance of different transmission routes in a variety of hotspots, for example, in restaurants~\cite{lu2020covid}, call centers~\cite{parkearly},  choirs~\cite{hamner2020high}, and churches~\cite{arkansas.church}. More of these real-world examples will allow policymakers to evaluate the effectiveness of different control measures -- for example, reducing crowd density, fever screening, rearranging seating layout, requiring masks, good ventilation -- in eliminating major transmission routes in confined spaces.

\section*{Stochasticity, superspreading events and the future of SARS-CoV-2}

Multiple lines of evidence at the individual- and population-level strongly indicate the role of SSEs in the transmission dynamics of SARS-CoV-2 and that we should not overlook the heterogeneity in numbers of secondary infections~\cite{friedenearly}. Our mental picture should not be that most people transmit to two or three other people, but instead a small number of infections dominate the transmission while most others fail to have secondary infections. The distribution of \R is over-dispersed with a high probability of extinction on the lower end, and a long tail on the higher end.

Outbreaks will be less likely to take off because of the high probability of extinction. At the early stage we will see more randomness and stochasticity, and it seems more explosive with huge number of cases reported in the first few generations. But once it takes off, it still becomes a stable exponential as per classic deterministic models. 
Behavioral change, mild to moderate non-pharmaceutical interventions to lower population-wide \Reff, as well as other factors such as population density~\cite{bialek2020geographic} and climate~\cite{oreilly2020effective} might be more impactful, as a lower population \Reff increases the extinction probability and increases the success probability of `cutting the tail'. Therefore, all measures that could potentially reduce \R should be considered and implemented, even if some only have a minor impact.

While SSEs are fueling this outbreak, we have an opportunity to take advantage of this heterogeneity in transmission, and use it to risk-stratify populations and locations for public health interventions and interrupt future SSEs. Novel methods are needed to quickly predict, identify, or isolate individuals / hotspots with the potential for causing SSEs.
It may prove difficult to identify and isolate individuals with the potential for causing SSEs and infectious individuals still transmit. This is compounded with the sizeable proportion of pre-symptomatic, as well as asymptomatic or minimally symptomatic individuals who can actively transmit~\cite{althouse2015asymptomatic,inui2020chest, hu2020clinical,nishiura2020estimation}.
If SSE predictors cannot be identified, then every infectious individual has the opportunity to cause a SSE if exposed to sizable susceptible populations. Therefore, it is crucial to understand types of hotspots and patterns of transmission for each type, as interventions might have to focus on all hotspots at high risk for SSEs and limiting gatherings at these places, and/or through rapid-and-extensive testing and contact tracing (both traditional and digital) to identify  pre-symptomatic and asymptomatic people~\cite{ferretti2020quantifying}. Contact tracing efforts should have an explicit goal to understand types of transmission and hotspots, so that the characterization of transmission could be used to adapt and prioritize other recommendations such as masks and mass gatherings. Further, it thus remains important to be cautious in reopening populations undergoing {\em cordons sanitaires} until transmission routes in different types of hotspots are well understood, or when safe and effective COVID-19 treatments and vaccines are available.

\section*{Acknowledgements}
The authors thank Philip Welkhoff, Mike Famulare, Anna Bershteyn, and Lowell Wood for insightful discussions and comments. B.M.A. and E.A.W. are supported by Bill and Melinda Gates through the Global Good Fund. 
J.C.M. is supported by startup funding from La Trobe University.
S.V.S. is supported by startup funds provided by Northeastern University. 
A.A. acknowledges financial support from the Sentinelle Nord initiative of the Canada First Research Excellence Fund and from the Natural Sciences and Engineering Research Council of Canada (project 2019-05183).
L.H.-D. acknowledges support from the National Institutes of Health 1P20 GM125498-01 Centers of Biomedical Research Excellence Award.
The funders had no role in study design, data collection, data analysis, the decision to publish, or preparation of the manuscript.

\clearpage
\pagebreak
\small
\bibliographystyle{science}

\bibliography{sample}

\end{document}